\documentclass[12pt]{article}

\usepackage{graphicx}
\usepackage{epstopdf}
\usepackage[centertags]{amsmath}
\usepackage{amssymb}
\usepackage{amsthm}
\usepackage{amsfonts}
\usepackage{ccaption}
\usepackage[usenames]{color}
\usepackage{mathrsfs}
\usepackage[textsize=tiny, textwidth=2cm,color=red,backgroundcolor=pink]{todonotes}
\usepackage[
      colorlinks=true,
      linkcolor=blue,  
      urlcolor=blue,    
      filecolor=black,     
      citecolor=red,
      pdfstartview=FitV,
      pdftitle={},
        pdfauthor={Veronika Hubeny},
        pdfsubject={},
        pdfkeywords={},
        pdfpagemode=None,
        bookmarksopen=true    
      ]{hyperref}
       
\usepackage{rotating}

\makeatletter
\@addtoreset{equation}{section}

\makeatletter
\renewcommand\section{\@startsection {section}{1}{\z@}%
                                   {-3.5ex \@plus -1ex \@minus -.2ex}
                                   {2.3ex \@plus.2ex}%
                                   {\normalfont\large\bfseries}}
\renewcommand\subsection{\@startsection{subsection}{2}{\z@}%
                                     {-3.25ex\@plus -1ex \@minus -.2ex}%
                                     {1.5ex \@plus .2ex}%
                                     {\normalfont\bfseries}}

  \captionnamefont{\bfseries}
  \captiontitlefont{\small\sffamily}
  \captiondelim{: }
  \hangcaption

\def\sec#1{\S\ref{#1}}
\def\fig#1{Fig.\,\ref{#1}}
\def\req#1{(\ref{#1})}

\def\CH{{\cal H}}

\def\CS{{\cal S}}

\def\AdS#1{AdS$_{#1}$}
\def\SAdS#1{Schwarzschild-AdS$_{#1}$}
\def\RAdS{R_{\rm AdS}}

\title{{\bf \Large The Fluid/Gravity Correspondence:}\\
{\bf \large a new perspective on the Membrane Paradigm}}

\author{\normalsize 
Veronika E. Hubeny\footnote{veronika.hubeny@durham.ac.uk},  \\
\small \sl   Centre for Particle Theory \& Department of
Mathematical Sciences,
\\[-1.5mm]
\small \sl Science Laboratories, South Road, Durham DH1 3LE, United Kingdom. \\
}

\begin{document}

\setlength{\baselineskip}{16pt}
\begin{titlepage}
\maketitle
\begin{picture}(0,0)(0,0)
\put(360, 240){DCPT-10/65} 
\end{picture}
\vspace{-36pt}

\begin{abstract}
This talk gives an overview of the recently-formulated Fluid/Gravity correspondence, which was developed in the context of gauge/gravity duality.  Mathematically, it posits that Einstein's equations (with negative cosmological constant) in $d+1$ dimensions capture the (generalized) Navier-Stokes' equations in $d$ dimensions.  Given an arbitrary fluid dynamical solution, we can systematically construct a corresponding asymptotically AdS black hole spacetime with a regular horizon whose properties mimic that of the fluid flow.  Apart from an overview of this construction, we describe some of its applications and implications.
 \end{abstract}
\thispagestyle{empty}
\setcounter{page}{0}
\end{titlepage}

\renewcommand{\thefootnote}{\arabic{footnote}}


\section{Introduction}
\label{s:intro}

Instead of starting by indicating what the title of my talk, {\it The Fluid/Gravity Correspondence}, is,  I will start by mentioning what it is not.
For several decades now, relativists have been intrigued by the idea that spacetime, or some important part of it like black hole horizons, might resemble a fluid.
Already in the 70's, black hole thermodynamics \cite{Bekenstein:1973ur,Hawking:1974sw,Bardeen:1973gs} laid its foundations with the spectacular realization that stationary black hole horizons have thermodynamic properties such as temperature and entropy, much like fluids; in fact the generalized 2nd law of thermodynamics treats black hole entropy on par with external matter entropy.  
In the early 80's, analog models of  black holes \cite{Unruh:1980cg} illustrated the converse notion, that fluids can admit sonic horizons and even the analog of Hawking temperature; indeed they can reproduce the kinematic aspects of black holes.
I should also mention that in many respects, black holes actually do exhibit behaviour similar to liquid droplets.  For example, recently \cite{Cardoso:2006ks} 
used fluid analog models to study higher dimensional black string Gregory-Laflamme instability \cite{Gregory:1993vy} as a Rayleigh-Plateau instability of liquid droplets \cite{Plateau:1873fk,Rayleigh:1878uq}, and \cite{Rezzolla:2010df} observed fluid-like recoil behaviour of horizon in studying anti-kick of merging black holes.
Perhaps most famously, the black hole Membrane Paradigm \cite{Thorne:1986iy,Damour:1978cg} developed in the mid-80s, realizes the idea that for external observers, black holes behave much like a fluid membrane, endowed with physical properties such as viscosity, conductivity, etc..  In particular, the dynamics of this membrane is described by the familiar laws of fluid dynamics, namely the Navier-Stokes equations, supplemented by Ohm's law and so forth.

All of these ideas contain the element of black holes sharing certain fluid properties.
However, the Fluid/Gravity correspondence, which is the subject of this talk, is none of these.


So let me now preview what the Fluid/Gravity correspondence is.
It is a relation between fluid dynamics on a fixed $(3+1)$-dimensional background, and gravity (specifically Einstein's general relativity with negative cosmological constant) in $4+1$ dimensions.
Mathematically, it posits that Einstein's equations (with negative cosmological constant) in $d+1$ dimensions capture the (generalized) Navier-Stokes equations in $d$ dimensions.  Given an arbitrary fluid dynamical solution, we can systematically construct a corresponding asymptotically AdS black hole spacetime with a regular horizon whose evolution mimics that of the fluid flow. 
The specific correspondence was formulated within the context of the gauge/gravity duality just a few years ago by 
Bhattacharyya, Minwalla, Rangamani, and myself in \cite{Bhattacharyya:2008jc}, building on previous works  \cite{Policastro:2001yc,Janik:2005zt,Bhattacharyya:2007vs}
and since then has been generalized and applied in hundreds of further works.\footnote{
For recent reviews, see e.g.\ \cite{Rangamani:2009xk,Hubeny:2010fk}, and in a broader context of time-dependence in AdS/CFT, \cite{Hubeny:2010ry}.}  


As with most ideas which bridge several fields, there are many potential applications and opportunities for cross-fertilization between the fields. 
We saw an example of this in Gary Horowitz's talk, where gravitational calculations provided insight into certain condensed matter systems. 
Broadly speaking, the fluid/gravity correspondence has applications not only to black hole physics, but also to strongly coupled field theories, as well as fluid dynamics itself.  
Since a given fluid solution specifies a corresponding evolving and non-uniform black hole solution (to arbitrary accuracy in the long-wavelength regime), the fluid in effect provides a useful window into {\it generic} black hole dynamics, no longer constrained by any symmetries.
Conversely, we can use the gravity side to learn about the characteristic properties of the gauge theory plasma, such as transport coefficients of the conformal fluid.  Such quantities depend on the underlying microscopic structure and are notoriously difficult to calculate directly on the field theory side; nevertheless within our framework, gravity actually {\it determines} them. This is in fact useful even for experimental physics, since such conformal fluid to a degree mimics the physics of the quark-gluon plasma currently observed at the Relativistic Heavy Ion Collider, as well as that of certain condensed matter systems.\footnote{
For nice reviews, see e.g.\ \cite{Mateos:2007ay, Gubser:2009md,McGreevy:2009xe,Hartnoll:2009sz}.}
Finally,  since the low-energy effective description of gauge theory is fluid dynamics, the fluid/gravity correspondence suggests intriguing applications to hydrodynamics as such.  Despite decades of theoretical as well as numerical, observational, and experimental study of hydrodynamics, there are still many deep questions which remain to be answered.    For example, one of the famous Clay Millennium Prize Problems concerns the global regularity (existence and smoothness) of the Navier-Stokes equations \cite{Fefferman:2000fk}. Intriguingly, the solutions often include turbulence, which, in spite of its practical importance in science and engineering, still remains one of the great unsolved problems in physics.
The fluid/gravity framework allows us to `geometrize' the set-up, thereby providing a new perspective on these long-standing hydrodynamical puzzles.


The plan for the rest of the talk is the following. 
I will first briefly present the essential background, recalling the highlights from the gauge/gravity duality.  I will then describe our starting point, namely the correspondence for the configuration describing a global equilibrium.  Considering deformations of this `seed' configuration, we will be able to include the important physics of dissipation and to construct genuinely time-dependent solutions.  I will describe the method of obtaining these solutions in a `boundary-derivative' expansion, first at the conceptual level and then more formally.  Having indicated how to obtain a generic solution to arbitrary order  in this expansion, I will discuss the solution to second order,  obtained in \cite{Bhattacharyya:2008jc,Bhattacharyya:2008xc,Bhattacharyya:2008mz}.
In particular, I will focus on identifying the event horizon in the bulk geometry and extracting the transport  coefficients in the boundary fluid.
Finally, I will briefly mention some important generalizations of the framework and discuss further applications of the fluid/gravity correspondence.

\section{Background: gauge/gravity duality}
\label{s:gaugegrav}


Underlying the fluid/gravity framework is the gauge/gravity (or AdS/CFT) duality.
In a nutshell, this duality \cite{Maldacena:1997re, Gubser:1998bc, Witten:1998qj}\footnote{
The AdS/CFT correspondence is  comprehensively reviewed in the classic reviews \cite{Aharony:1999ti,DHoker:2002aw}.  For more recent reviews see e.g.\ \cite{McGreevy:2009xe,Polchinski:2010hw}.}
 relates a particular strongly coupled non-abelian gauge theory in $d$ dimensions to string theory, which in certain regime reduces to classical gravity, on $(d+1)$-dimensional asymptotically Anti de Sitter (AdS)  spacetime.
 
It is worth noting the key aspects of this correspondence.
Most conspicuously, the gauge/gravity duality relates a gravitational theory to non-gravitational one.  In fact,  the gauge theory in a sense provides a formulation of quantum gravity on asymptotically AdS spacetime.  This has fueled a large amount of research during the last decade, as one hopes to solve many long-standing quantum gravitational problems by recasting them in a non-gravitational language.
More intriguingly, the correspondence is {\it holographic}: the two dual theories live in different number of dimensions.\footnote{
In fact, the holographic principle \cite{tHooft:1993gx,Susskind:1994vu}, motivated by the peculiar non-extensive nature of black hole entropy, was proposed already prior to the AdS/CFT correspondence, but its best-understood realization appears in the AdS/CFT context.
} 
A useful conceptualization of the duality is to think of the gauge theory as `living on the boundary' of AdS.
We therefore refer to the gravity side as the ``bulk" and the gauge side as the ``boundary" theory.
Finally, AdS/CFT constitutes a strong/weak coupling duality; the strongly-coupled field theory can be accessed via the semi-classical gravitational dual, which has obvious computational as well as conceptual advantages.  Hence the information flow, namely using one side of the duality to learn about the other, proceeds fruitfully in both directions.


Let me now describe several more specific features of the correspondence.
Distinct asymptotically AdS (bulk) geometries correspond to distinct states in the (boundary) gauge theory.
The pure AdS bulk geometry, i.e.\ the maximally symmetric negatively curved spacetime, corresponds to the vacuum state of the gauge theory.
Deforming the bulk geometry (while maintaining the AdS asymptotics) corresponds to exciting the state (within the same theory).
Specifically, such metric perturbations are related to the stress(-energy-momentum) tensor expectation value in the CFT.   More importantly in the present context, a large\footnote{
AdS is a space of constant negative curvature, which introduces a length scale, called the AdS scale $\RAdS$, corresponding to the radius of curvature.  The black hole size is then measured in terms of this AdS scale; large black holes have horizon radius $r_+  > \RAdS$.  Here we will take the large black hole limit $r_+  \gg \RAdS$, and therefore consider so-called {\it planar} Schwarzschild-AdS black holes.
} Schwarzschild-AdS black hole corresponds to a thermal state in the gauge theory.  This can be easily conceptualized as the late-time configuration a generic state evolves to: in the bulk, the combined effect of gravity and negative curvature tends to make a generic configuration collapse to form a black hole which settles down to the Schwarzschild-AdS geometry, while in the field theory, a generic excitation will eventually thermalize. 
Note that although the underlying theory is supersymmetric, the correspondence applies robustly to non-supersymmetric states such as the black holes mentioned above.  In this sense, supersymmetry is {\it not} needed for the correspondence.

On the boundary, the essential physical properties of the gauge theory state (such as local energy density, pressure, temperature, entropy current, etc.) are captured by the boundary stress tensor, which in turn is induced by the bulk geometry and can be extracted via a well-defined Brown-York type procedure \cite{Balasubramanian:1999re}.\footnote{
For example, for asymptotically \AdS{d+1} spacetimes, the prescription of \cite{Balasubramanian:1999re} gives
$$
T^{\mu\nu} = \lim_{\Lambda_\text{c} \to \infty}\; \frac{\Lambda_\text{c}^{d-2}}{16\pi \, G_N} \, \left[ K^{\mu\nu} - K \, \gamma^{\mu\nu} - (d-1)\, \gamma^{\mu\nu} - \frac{1}{d-2}\,  \left(R^{\mu\nu} -\frac{1}{2}\, R \, \gamma^{\mu\nu}\right)\right] 
$$
where $\gamma^{\mu\nu}$ is the $d$-dimensional metric induced on a $r = \Lambda_\text{c}$ cutoff surface, $R^{\mu\nu}$ and $R$ are the corresponding Ricci tensor and scalar, $K^{\mu\nu}$ and $K$ are the extrinsic curvature and its trace, and $G_N$ is the Newton's constant in $d+1$ dimensions.  See also \cite{deHaro:2000xn}.}
It is important to distinguish the two stress tensors one might naturally consider.  In our framework, the {\it bulk} stress tensor appearing on the RHS of the bulk  Einstein's equation is set to zero, so that the bulk solutions $g_{ab}$ correspond to general vacuum black holes with negative cosmological constant but no other matter content.  On the other hand, the {\it boundary} stress tensor $T^{\mu\nu}$ is non-zero; it captures the matter content of the boundary theory, its conservation determines the dynamics, but it does not curve the boundary spacetime \`a la Einstein's equations since the boundary metric is non-dynamical and fixed (in our case to the 4-dimensional Minkowski spacetime).

To summarize,\footnote{
We use the following notation for the coordinates:  
 the bulk line element $ds^2 = g_{ab} \, dX^a \, dX^b$ depends on all bulk directions $X^a = (r, x^\mu)$ which consist of the radial direction $r$ and the `boundary' spacetime directions $x^{\mu} = (t, x^i)$.
The $d+1$ dimensional bulk action is given by
$$ \CS_{\rm bulk} = \frac{1}{16 \pi \, G_N} \, \int d^{d+1} X \, 
\sqrt{-g} \, (R - 2 \,  \Lambda) \ . $$
} the boundary fluid is specified by the boundary stress tensor $T^{\mu\nu}(x^\mu)$, while the bulk geometry is specified by the bulk metric $g_{ab}(r,x^\mu)$.
The bulk dynamics is determined by EinsteinÕs equations,
\begin{equation}
 E_{ab} \equiv R_{ab} - \frac{1}{2} \, R \, g_{ab} + \Lambda  \, g_{ab} =0 \ ,
 \label{Eeq}
 \end{equation}
while the boundary dynamics is determined by stress tensor conservation,
\begin{equation}
\nabla_\mu T^{\mu\nu} = 0 \ .
\label{Tcons}
\end{equation}	
In the following, we'll see that \req{Tcons}  actually arises from \req{Eeq}; in this sense, bulk gravity gives rise the boundary fluid dynamics.

\section{Global equilibrium}
\label{s:globeq}


Let us now examine the planar Schwarzschild-AdS black hole which, as already mentioned, describes a state in global thermal equilibrium.
The metric of the planar \SAdS{5} black hole is 
\begin{equation}
ds^2 = r^2 \, \left( -f(r) \, dt^2 + \sum_{i=1}^3 (dx^i)^2 \right) + {dr^2 \over r^2 \, f(r)} \ , \qquad {\rm where} \ \ 
f(r) \equiv 1- \frac{r_+^4}{ r^4} \ .
\label{SAdSmet}
\end{equation}	
This spacetime has a spacelike curvature singularity at $r=0$, cloaked by a regular event horizon at $r=r_+$, and a timelike boundary at $r= \infty$. 
\begin{figure}
\begin{center}
\includegraphics[width=2in]{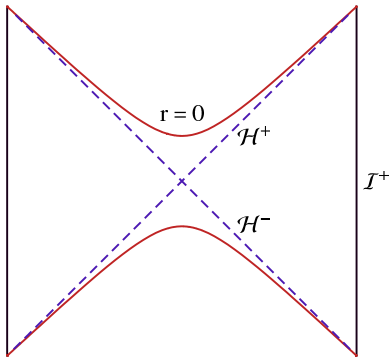}
\caption{Penrose diagram for the planar Schwarzschild-Ads$_5$ black hole given by \req{SAdSmet}.  The top and bottom (red) curves correspond to the curvature singularity, the diagonal dashed (blue) lines to the horizon, and the vertical (black) lines to the AdS boundaries.}
\label{f:SAdSPD}
\end{center}
\end{figure}
The causal structure of this solution is described by the Penrose diagram of \fig{f:SAdSPD}.  
An important quantity is the temperature of this black hole, determined from 
the  surface gravity (with respect to $(\frac{\partial}{\partial t})^a$ at $r= r_+$) to be
\begin{equation}
T = \frac{r_+}{\pi} \ .
\label{planarbhtemp}
\end{equation}
Note that unlike the usual asymptotically flat Schwarzschild black hole where the temperature scales inversely with the black hole size, here it scales linearly; this confirms that such a black hole is thermodynamically stable, as required of thermal equilibrium.

This solution is static and translationally invariant in the boundary spatial directions $x^i$. One can in fact generate a $4$-parameter family of solutions by scaling $ r$ and boosting in $\mathbb{R}^{3,1}$ with normalized $4$-velocity $u^\mu$.   Moreover, we can pass to the analog of ingoing Eddington-Finkelstein coordinates, so as to render the metric manifestly regular on the horizon.  This allows us to express the planar Schwarzschild-AdS$_5$ black hole \req{SAdSmet} in more convenient (regular and boundary-covariant) coordinates 
\begin{equation}
ds^2 =-2\, u_{\mu}\, dx^{\mu} \,  dr 
+ r^2\, \left( \eta_{\mu\nu} + \frac{\pi^4 \, T^4}{r^4} \, u_{\mu}\, u_{\nu} \right) \, dx^{\mu} \,  dx^
{\nu}  \ , 
\label{SAdSzero}
\end{equation}
parameterized by the black hole temperature $T$ and the horizon velocity $u^{\mu}$.  Note that since $u_{\mu} \, u^{\mu}= -1$, \req{SAdSzero} constitutes a 4-parameter family of solutions of \req{Eeq}, describing stationary black holes.

Let us now turn to the boundary description of such a state.  The boundary stress tensor induced by the bulk metric \req{SAdSzero} is (upon setting 
$\frac{1}{16\pi \, G_N} = 1$)
\begin{equation}
T^{\mu \nu} =  \pi^4 \, T^4 \, \left(\eta^{\mu \nu} + 4 \, u^\mu \, u^\nu \right) \ .
\label{TzeroO}
\end{equation}
This describes a perfect fluid at temperature $T$, moving with velocity $u^{\mu}$ on the flat 4-dimensional background $\eta_{\mu\nu}$.
Note that this stress tensor is traceless, $T_{\ \mu}^{\mu} =0$, as befits a conformal fluid. More importantly, there is no dissipation in the system.\footnote{
This is manifest for stationary fluid, but even if $T$ and $u^\mu$ vary in time, the perfect fluid form of the stress tensor \req{TzeroO} disallows dissipation, as can be verified from the vanishing divergence of corresponding entropy current.}
In order to describe more general time-dependent and dissipative systems, we need to go beyond a perfect fluid in global equilibrium.

\section{Nonlinear deformations away from global equilibrium}
\label{s:nonlin}

We will now motivate how to go about constructing such a general set of solutions.  We first focus on the stress tensor, explaining how its form is determined by the symmetries of the set-up, leaving us with a finite number of undetermined coefficients.   We then recall how some of these coefficients have been previously obtained from linearized analysis in the gravity dual.  Finally, we explain how to go beyond the linearized regime to construct our generic solutions in the bulk.

\subsection{Fluids with dissipation}
\label{s:dissip}


Dissipation is a crucial aspect of the physics, allowing the state to settle down at late times.
If the stress tensor is to capture dissipation, it must allow for variations of  $T$ and $u^{\mu}$.  However, in order 
to have a sensible fluid description, these variations are constrained to lie in the  so-called {\it long wavelength regime}: the scale of variation $L$ of the fluid variables  $T$ and $u^{\mu}$ must be large compared to the microscopic scale $1/T$ -- otherwise these thermodynamic variables would be meaningless.  
This automatically provides a small parameter 
\begin{equation}
\epsilon \equiv \frac{1}{L \, T} \ll 1 \ ,
\label{epsilon}
\end{equation}	
and naturally allows us to expand the stress tensor $T^{\mu\nu}$ in `boundary derivatives' $\partial_\mu(\ldots)$.  In such an expansion, terms of order $(\partial_\mu u_\nu)^n , \ldots , \partial_\mu^n  \, u_\nu$ will be suppressed by $\epsilon^n$. 
In particular, we can expand the stress tensor as
\begin{equation}
T^{\mu \nu} =  \pi^4 \, T^4 \, \left(\eta^{\mu \nu} + 4 \, u^\mu \, u^\nu \right) + \Pi_{(1)}^{\mu \nu}+ \Pi_{(2)}^{\mu \nu} + \ldots \ ,
\label{TtwoO}
\end{equation}
where $\Pi_{(1)}^{\mu \nu}$ contains dissipative terms composed of single-derivative expressions such as $\partial^\mu  u^\nu$, the next term $\Pi_{(2)}^{\mu \nu}$ contains the second order dissipative terms, and so on.
As mentioned above, the dynamics is determined by the conservation equations \req{Tcons}, 
which become more complicated as one includes more terms  in $T^{\mu \nu}$.  For the zeroth-order $T^{\mu \nu}$ given by the perfect fluid \req{TzeroO}, this yields mass conservation and Euler equation; when one includes dissipation, the stress tensor conservation is described by the generalized\footnote{
There are two generalizations to the form described in conventional (non-relativistic) fluid dynamics \cite{Landau:1965uq}: one arises from including terms beyond first order in boundary derivatives, and another from the fact that our fluid is relativistic, with pressure comparable to the energy density.  
} Navier-Stokes equations.

It turns out that \req{TtwoO} is a very useful way to package the stress tensor.  At each order, the form of the stress tensor is actually determined by symmetries, leaving just a finite number of undetermined ÔtransportÕ coefficients.  Since we are dealing with a conformal fluid, the stress tensor has to be Weyl covariant, as well as generally covariant in the boundary directions.  This procedure of using the Weyl-covariant formalism \cite{Loganayagam:2008is} is so robust that we can equally easily write the form of a more general $d$-dimensional dissipative stress tensor for a conformal fluid living on a fixed background with metric $\gamma_{\mu\nu}$, to second order:
\begin{equation}
\begin{split}
T^{\mu\nu} =&\ P \, \left(\gamma^{\mu\nu}+d  \, u^\mu  \, u^\nu \right) 
-2 \, \eta \, \sigma^{\mu\nu}\\
&+2 \, \eta  \,  \left[\tau_1 \,  u^{\lambda} \, \mathcal{D}_{\lambda}\sigma^{\mu \nu} -\tau_\epsilon \, (\omega^{\mu}{}_{\lambda} \, \sigma^{\lambda \nu}+\omega^\nu{}_\lambda \,  \sigma^{\lambda\mu}) \right]
+ \xi_C \, C^{\mu\alpha\nu\beta} \, u_\alpha  \, u_\beta \\ 
&+ \xi_\sigma \, [ \sigma^{\mu}_{\ \lambda} \,\sigma^{\lambda \nu}
- \frac{ P^{\mu \nu}}{d-1} \sigma_{\alpha \beta} \, \sigma^{\alpha \beta}]
+\xi_\omega \, [ \omega^{\mu}_{\ \lambda} \,\omega^{\lambda \nu}
+ \frac{ P^{\mu \nu}}{d-1} \omega_{\alpha \beta} \, \omega^{\alpha \beta}] \ ,
\label{ddim2T}
\end{split}
\end{equation}
where $P$ is the pressure and we have used various standard quantities built out of the velocity $u^\mu$ and the background metric $\gamma_{\mu\nu}$; in particular, $\sigma_{\mu\nu}$ and $\omega_{\mu\nu}$ are the shear and the vorticity of the fluid, respectively,  $P^{\mu\nu} = \gamma^{\mu\nu} + u^\mu  \, u^\nu $ is the spatial projector, $\mathcal{D}_{\lambda}$ is the Weyl-covariant derivative, and $C_{\mu\nu\alpha\beta}$ is the Weyl tensor for $\gamma_{\mu\nu}$.  In the above expression, the 0th and 1st order terms appear on the first line, whereas the 2nd order terms fill the remaining two lines.

\subsection{Transport coefficients from linearized gravity}
\label{s:transpcoeff}

The shear viscosity  $\eta$ and the five second-order transport coefficients, $\tau_1$, $\tau_\epsilon$, $\xi_C$, $\xi_\sigma$, and $\xi_\omega$, are not determined from the symmetries.
These transport coefficients depend on the microscopic structure of the fluid; they could be in principle measured, or calculated from first principles.
However, both of these approaches are rather difficult, since the gauge theory is strongly coupled.
Nevertheless, as we will shortly see, the bulk dual in fact determines these transport coefficients uniquely.
Although this will occur very naturally within the fluid/gravity framework, one should note that the transport coefficients can already be extracted in the linearized regime, from quasinormal modes\footnote{
These modes describe small fluctuations of a black hole, namely its ringing and settling down.  Mathematically, they are related to the poles of the retarded Green's function.
A good pre-AdS/CFT review is \cite{Kokkotas:1999bd}, in AdS/CFT context these were first discussed in \cite{Horowitz:1999jd}, and a recent extensive review of quasinormal modes in context relevant to the present set-up appears in \cite{Berti:2009kk}.
}
of the planar black hole.


To see how this works in more detail, let us first recall that  the black hole quasinormal modes encode the field theory's return to thermal equilibrium \cite{Horowitz:1999jd}.
Most modes decay with a characteristic timescale related to the size of the black hole $r_+$, but there are also so-called hydrodynamic modes which can have arbitrarily long wavelength and small frequency, and therefore fall within the long-wavelength regime discussed above.
Such modes with hydrodynamic dispersion relations were first considered in \cite{Policastro:2002se,Policastro:2002tn} and describe a propagating sound mode with linear dispersion and shear mode with damped quadratic dispersion.\footnote{
Extracting linearized hydrodynamics from linearized gravity has been pursued vigorously over the years; for a nice review, see \cite{Son:2007vk}.  For the dispersion relation describing the sound and shear modes of AdS black holes, see e.g.\ Fig.7 and Fig.4 of \cite{Morgan:2009pn}, respectively.
}
One can then use linear response theory to compute the  transport coefficients. This analysis not only confirmed the relation between classical dynamics in a black hole background and the physics of a strongly coupled plasma, but it also prompted the famous bound \cite{Kovtun:2004de} on the ratio of shear viscosity to entropy density, $\eta/s \ge \frac{1}{4\pi}$.  This bound is saturated by a large class of two-derivative theories of gravity, and it is indeed experimentally satisfied by all presently-known systems in nature. Intriguingly, cold atoms at unitarity and quark-gluon plasma both come near to saturating the bound \cite{Schafer:2009dj}.


\subsection{Constructing a generic black hole geometry}
\label{s:bulkconstr}

Rather than restricting attention to linearized gravity around a fixed black hole background, we now turn to the main task of finding a bulk solution of the full Einstein's equations, capable of describing arbitrarily large deviations from the stationary planar black hole \req{SAdSzero} in the long-wavelength regime.
Of course, solving the full Einstein's equations for a generic ansatze is prohibitively difficult, but we will see that the long-wavelength regime renders the problem tractable.  Before explaining the actual construction, we first provide a conceptual motivation for the method.

\begin{figure}
\begin{center}
\includegraphics[width=2.5in]{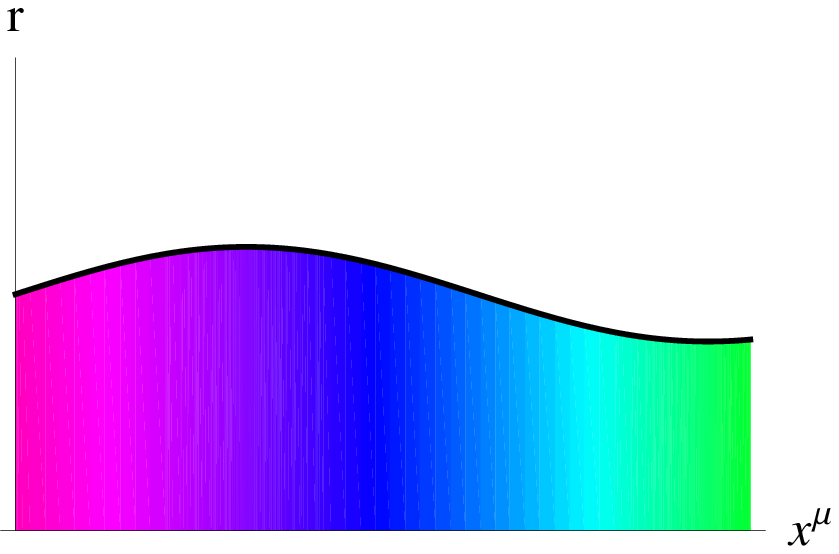}
\hspace{1cm}
\includegraphics[width=2.5in]{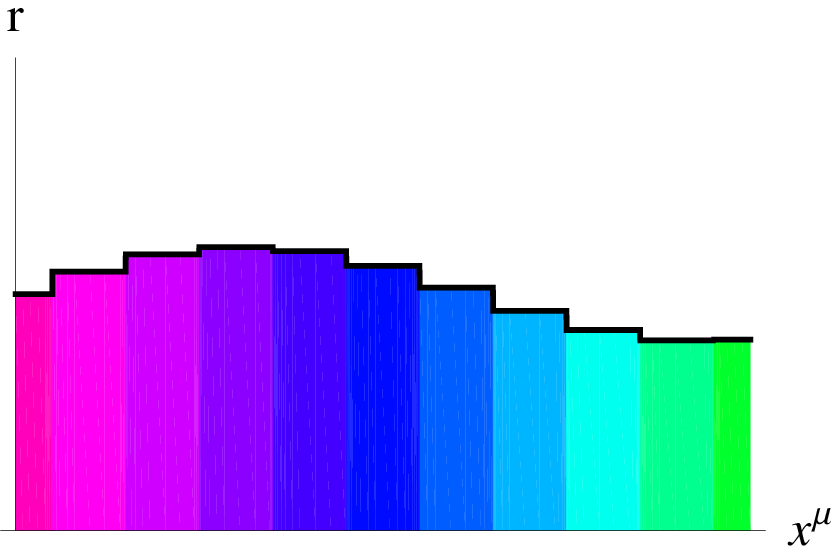}
\caption{Cartoon of `tubewise approximation' of slowly-varying configuration by a corresponding piecewise-constant one.}
\label{f:contdiscr}
\end{center}
\end{figure}

Let us suppose that the `parameters' $T$ and $u^\mu$ describing the black hole in \req{SAdSzero} vary slowly in $x^\mu$.
Then at each $x^\mu_0$, the geometry should look approximately like a
black hole with temperature $T(x_0)$ and velocity $u^\mu(x_0)$.  We refer to the bulk spacetime region in the neighborhood of a fixed $x^\mu$ but extended over all $r$ as a `tube'; and we say that in the long-wavelength regime, the bulk geometry `tubewise' approximates a planar black hole with specific velocity and temperature.  
We illustrate this idea in the cartoon of \fig{f:contdiscr}, where the curve indicates the variation of the temperature  $T(x^\mu)$ and the color-coding the variation of (some component of) the velocity.  The slower such variations are, the better can we approximate the configuration with piecewise-constant tubes.
Our task then  is to patch such tubes together to construct a non-uniform and time-evolving black hole. 


Of course, if we just replace $u_\mu$ and $T$ in the metric (\ref{SAdSzero}) by $T(x)$ and $u^\mu(x)$, the resulting metric (call it $g^{(0)}_{ab}$) will no longer solve Einstein's equations  (\ref{Eeq}). 
However, it is a manifestly regular metric which approaches a solution in the limit of infinitely slow variations.
This enables us to use the metric $g^{(0)}_{ab}$  as a starting point for constructing an iterative solution.
The requirement of slow variations can be written schematically as
\begin{equation}
\frac{\partial_\mu  \log T}{T} \sim \mathcal{O}(\epsilon) \ , \qquad \frac{\partial_\mu u}{T} \sim \mathcal{O}(\epsilon) 
\label{}
\end{equation}	
where $\epsilon$ is a small parameter.  In terms of the fluid description, it is indeed the same parameter \req{epsilon}  (counting the number of $x^{\mu}$ derivatives) which ensured that the configuration is in local equilibrium and therefore describable as a fluid.
Using $\epsilon$ as an expansion parameter, we  expand the metric and the fields $u_\mu(x)$ and $T(x)$ as
\begin{equation}
 g_{ab} = \sum_{k=0}^\infty \, \epsilon^k \, g^{(k)}_{ab} \ , \qquad
 T =\sum_{k=0}^\infty \, \epsilon^k \, T^{(k)}  \ , \qquad
u_{\mu} =\sum_{k=0}^\infty \, \epsilon^k \, u_{\mu}^{(k)} \ . 
\label{expand}
\end{equation}
We can then substitute the expansion \req{expand} into Einstein's equations \req{Eeq}, and find the solution order by order in $\epsilon$.
The term $g^{(k)}_{ab}$ corrects the metric at the  $k^{{\rm th}}$ order, such that  Einstein's equations will be satisfied to ${\mathcal O}(\epsilon^k)$ provided the functions $T(x)$ and $u^\mu(x)$ obey a certain set of equations of motion, which turn out to be precisely the stress tensor conservation equations \req{Tcons} of boundary fluid dynamics at ${\mathcal O}(\epsilon^{k-1})$.
Hence the resulting corrected metric can be constructed systematically to any desired order.  Importantly, the expansion remains valid well inside the event horizon, which allows verification of the regularity of such a solution.

Let us examine the structure of the equations a bit more explicitly.  Einstein's equations \req{Eeq} split up into two kinds:
Constraint equations, $E_{r \mu} = 0$ which implement stress-tensor conservation (at one lower order), and Dynamical equations $ E_{\mu\nu} = 0$ and $E_{rr} =0$  which allow determination of $g^{(k)}$.
Schematically, the latter take a miraculously simple form:
\begin{equation}
{\mathbb H}\left[g^{(0)}(u^{(0)}_\mu, T^{(0)})\right] g^{(k)}= s_k  \ ,
\label{schemEeq}
\end{equation}
where ${\mathbb H}$ is a second-order linear differential operator in the variable $r$ alone and $s_k$ are regular source terms which are built out of $g^{(n)}$ with $n \le k-1$.
 Since $g^{(k)}(x^\mu ) $ is already of ${\mathcal O}(\epsilon^k)$, and since every boundary derivative appears with an additional power 
of $\epsilon $, ${\mathbb H}$ is an ultra-local operator in the field theory directions. 
 Moreover, at a given $x^{\mu}$, the precise form of this operator ${\mathbb H}$ depends only on the local values of $T$ and $u^\mu$ but not on their derivatives at $x^{\mu}$.  Furthermore, we have the same homogeneous operator ${\mathbb H}$ at every  order in perturbation theory. 
 This allows us to find an explicit solution of (\ref{schemEeq}) systematically at any order.
The source term $s_k$ however gets more complicated with each order, and reflects the nonlinear nature of the theory. 
We solve the dynamical equations
$$g^{(k)} = {\rm particular}(s_k) + {\rm homogeneous}({\mathbb H})$$ 
subject to regularity in the interior and asymptotically AdS boundary conditions. 
The solution is guaranteed to exist,\footnote{
Using the rotational symmetry group of the seed solution (\ref{SAdSzero}) it turns out to be possible to make a judicious choice of variables such that the operator ${\mathbb H}$ is converted into a decoupled system of first order differential operators. It is then simple to solve the equation (\ref{schemEeq}) for an arbitrary source $s_k$ by direct integration.
For the details of the procedure, as well as discussion of convenient gauge choice,
 etc.,  see the original work \cite{Bhattacharyya:2008jc}  or the review \cite{Rangamani:2009xk}.}
 provided the constraint equations are solved.


Before turning to the solution itself, let us summarize the key points of our construction.
The iterative construction can in principle be
systematically implemented to arbitrary order in $\epsilon $
(which obtains correspondingly accurate solution).
The resulting black hole spacetimes actually correspond to not just a single solution or even a finite family of solutions, but rather a continuously-infinite set of (approximate) solutions, specified by four functions, $T(x)$ and $u^{\mu}(x)$, of four variables.  
The flip side of the coin is that while very general, such a metric is not fully explicit:  in order to be so, we need to use a given solution to fluid dynamics, which relates the functions $T(x)$ and $u^{\mu}(x)$, as input.
Nevertheless, given any such solution, the construction guarantees that the bulk geometry describes a black hole with regular event horizon.

\section{General solution}
\label{s:soln}


The solution for the bulk metric $g_{ab}(r,x^\mu)$ and the boundary stress tensor $T^{\mu\nu}(x^\mu)$ (written in terms of the temperature and velocity fields $T(x^\mu)$ and $u^\nu(x^\mu)$) was explicitly constructed to second order in the boundary derivative expansion in \cite{Bhattacharyya:2008jc}.  This solution was further studied in \cite{Bhattacharyya:2008xc}, where its regularity was confirmed by identifying the event horizon.
This construction was subsequently generalized to other contexts, as reviewed in \cite{Rangamani:2009xk}.
Since the solution for the second-order metric is page-long, here we only report the solution to first order for illustration.

To first order the bulk metric takes the form
\begin{eqnarray}
ds^2 &=&-2\, u_{\mu}\, dx^{\mu} dr 
+ r^2\, \left( \eta_{\mu\nu} + [1-f(r/\pi T)] \, u_{\mu}\, u_{\nu} \right) \, dx^{\mu}dx^{\nu} \nonumber \\
&+& 2r \left[ { r \over \pi T} \, F(r/\pi T)\, \sigma_{\mu\nu} +{1\over 3} \, u_{\mu}u_{\nu} \,\partial_{\lambda} u^{\lambda}  -  {1\over 2}\, u^{\lambda}\partial_{\lambda}\left(u_\nu u_{\mu}\right)\right] \, dx^{\mu} dx^{\nu} ,
\label{metfirstO}
\end{eqnarray}
where $f(r)$ is defined in \req{SAdSmet},  $F(r)$ is given by 
$$
F(r) \equiv \int_r^{\infty}\, dx \,\frac{x^2+x+1}{x (x+1) \left(x^2+1\right)} ={1\over 4}\, \left[\ln\left(\frac{(1+r)^2(1+r^2)}{r^4}\right) - 2\,\arctan(r) +\pi\right] ,
$$
$\sigma^{\mu\nu}= P^{\mu \alpha} P^{\nu \beta} \, 
\, \partial_{(\alpha} u_{\beta)}
-\frac{1}{3} \, P^{\mu \nu} \, \partial_\alpha u^\alpha $ is the shear,
and $T(x)$ and $u_\mu(x)$ are any slowly-varying functions which satisfy the conservation equation \req{Tcons} for the zeroth order perfect fluid stress tensor \req{TzeroO}.
Note that the first line of \req{metfirstO} corresponds to the zeroth order solution \req{SAdSzero}, whereas each of the terms in the second line have exactly one boundary derivative.\footnote{
Note that \req{metfirstO} does not have any $\partial_{\mu} T$ terms appearing explicitly, since by implementing the zeroth order stress tensor conservation, we have expressed the temperature derivatives in terms of the velocity derivatives.}

\begin{figure}
\begin{center}
\includegraphics[width=2.5in]{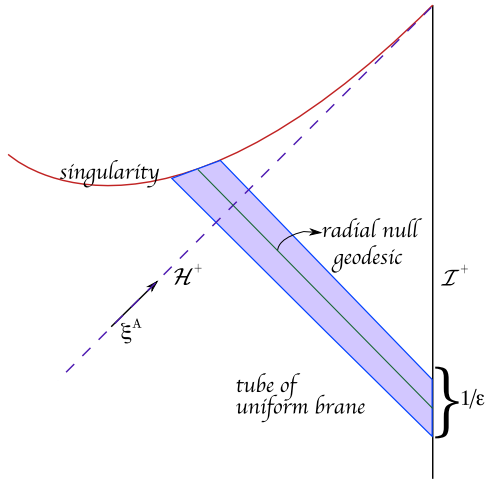}
\caption{The causal structure of the spacetimes dual to fluid mechanics illustrating the tube structure. The dashed line denotes the future event horizon $\CH^+$ generated by $\xi^A$, while the shaded tube indicates the region of spacetime over which the solution is well approximated by a tube of the uniform black brane.}
\label{f:tubes}
\end{center}
\end{figure}

As mentioned previously, this bulk solution is `tubewise' approximated by a planar black hole. This means that in each tube, defined by a small neighborhood of given $x^{\mu}$, but fully extended in the radial direction $r$, the radial dependence of the metric is approximately that of a boosted planar black hole at some temperature $T$ and horizon velocity $u^\mu$, with corrections suppressed by the rate of variation, $\epsilon$.  These parameters vary from one position $x^{\mu}$ to another in a manner consistent with fluid dynamics.   Our choice of coordinates is such that each tube extends along an ingoing radial null geodesic; see \fig{f:tubes}. 
Apart from technical advantages, this is conceptually rather pleasing, since it suggests a mapping between the boundary and the bulk which is natural from causality considerations. 
Physically, the solution \req{metfirstO} and its higher-order improvements of course describe a dynamically evolving black hole with infinitely extended but non-uniform event horizon.  The causal structure of this solution is preserved; in fact, dissipation will cause the black hole to approach a stationary solution \req{SAdSzero} at late times.


Let us now focus on the most salient feature of this geometry, namely its event horizon.
Assuming the dissipation causes our configuration to settle down to a stationary state at late times, we can find the event horizon as the unique null hypersurface with the correct late-time behavior.  
This can be solved algebraically, order-by-order in $\epsilon$, and takes the schematic form \cite{Bhattacharyya:2008xc}
\begin{equation}
r_+(x) = \pi \, T(x) + \frac{1}{ \pi \, T(x)} \, 
( \# \, \sigma_{\mu\nu}(x) \, \sigma^{\mu\nu}(x) +
 \# \, \omega_{\mu\nu}(x) \, \omega^{\mu\nu}(x) ) + \ldots
\label{}
\end{equation}	
Intriguingly, it turns out that within this derivative expansion, the location of the event horizon $ r_+(x^\mu)$ in the bulk is determined locally by the behavior of the temperature and velocity at a point $x^\mu$ (in particular it is insensitive to the metric at later times), rather than globally as usual in general relativity.  This curious locality is in fact allowed by the long wavelength regime,  wherein the horizon position varies sufficiently slowly.

\begin{figure}
\begin{center}
\includegraphics[width=2in]{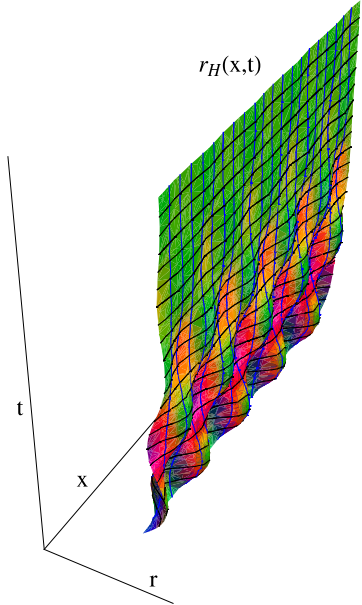}
\caption{A cartoon of the event horizon $r = r_+(x^\mu)$ sketched as a function of the time $t$ and one of the spatial coordinates $x^i$ (the other two spatial coordinates are suppressed).}
\label{f:wigglyhor}
\end{center}
\end{figure}

\fig{f:wigglyhor} gives a cartoon of the behaviour (for simplicity just the local temperature) of the event horizon for some generic fluid configuration.
We see that even if at early times (bottom of the sketch) the horizon is highly non-uniform, its evolution will tend to dissipate the inhomogeneities.  At late times (top part of the sketch), the horizon settles down to a stationary configuration.  Throughout, the evolution proceeds in such a way that the horizon area grows, as can be verified by explicit calculation.  This has an important consequence for the dual fluid dynamical description.  
The pull-back of the area form on the horizon provides a natural entropy current in the dual fluid.
Such entropy current then automatically satisfies the 2nd Law of thermodynamics.


Let us now turn to the induced fluid stress tensor on the boundary.  The stress tensor to first order in boundary derivatives can be easily obtained from  the bulk metric \req{metfirstO},  and takes the simple form
\begin{equation}
T^{\mu \nu} =\pi^4 \, T^4
\left( 4\,  u^\mu u^\nu +\eta^{\mu \nu}
\right)  - 2 \,\pi^3 \, T^3 \, \sigma^{\mu \nu} .
\label{TfirstO}
\end{equation}
Here the first two (derivative-free) terms describe a perfect fluid with pressure $\pi^4 \, T^4$, and correspondingly (using thermodynamics)  entropy density  $s=4 \, \pi^4 \, T^3$.  The shear viscosity $\eta$ of this fluid may be read off from the coefficient of $\sigma^{\mu\nu}$ and is given by $\pi^3 \, T^3$. 
Notice that this verifies that our system indeed saturates the famous bound \cite{Kovtun:2004de} on viscosity-to-entropy-density ratio 
$\frac{\eta}{s} = \frac{1}{4 \pi}$, in agreement with the well-known result of \cite{Policastro:2001yc}.


While we only verified previously-known results using the first-order stress tensor, at second order we can start making new predictions.
We have already written the form of the stress tensor in \req{ddim2T}, leaving only several coefficients to fix.  These transport coefficients are however important in determining the physical properties of the fluid.
For fluids in 3+1 dimensional flat spacetime, the second order transport coefficients were  obtained in \cite{Bhattacharyya:2008jc} (and concurrently by  \cite{Baier:2007ix}),
but here we will quote the more general result \cite{Bhattacharyya:2008mz,Haack:2008cp} pertaining to conformal fluids in $d$-dimensional curved spacetime $\gamma_{\mu\nu}$ with the stress tensor \req{ddim2T}:
\begin{equation}
\begin{aligned}
& P=\frac{1}{16\pi G_N} \, \left( \frac{4\pi \, T}{d} \right)^{d} \\
& \eta =\frac{s}{4\pi}=\frac{1}{16\pi G_N} 
    \, \left(\frac{4\pi \, T}{d} \right)^{d-1} \\
& \tau_1 = \frac{d}{4\pi \, T} \, \left( 1-\int_1^\infty dy \, 
    \frac{y^{d-2} -1 }{y \, (y^d -1)} \right) \\
& \tau_\epsilon = \frac{d}{4\pi \, T} \, \int_1^\infty dy \, 
    \frac{y^{d-2} -1 }{y \, (y^d -1)} \\
& \xi_\sigma = \xi_C = \frac{d}{4\pi \, T} \, 2\, \eta \\
& \xi_\omega = 0 \ .
\end{aligned} 
\label{transpgend}
\end{equation}	
Note that written suggestively in this way, we can discern intriguing relations between the coefficients, which hint at the specific nature of any conformal fluid which admits a gravitational dual.
For example, the results that $\xi_\sigma = \xi_C$ and $\xi_\omega = 0$ are universal but non-trivial from the fluid standpoint.  More intriguingly,
we see that\footnote{
in fact this relation continues to hold even for charged black holes   mentioned in \sec{s:concl} \cite{Haack:2008cp,Bhattacharyya:2008mz}.
} $\xi_\sigma = 2 \, \eta \, (\tau_1 + \tau_\epsilon)$  for all $d$.  

\section{Concluding remarks}
\label{s:concl}


The results summarized above have since been extended and generalized in a number of useful directions. 
As already indicated, one immediate set of generalizations involved relating a $d$-dimensional conformal fluid to asymptotically AdS$_{d+1}$ black hole (see \cite{VanRaamsdonk:2008fp} for the interesting case of $d=3$ and \cite{Haack:2008cp,Bhattacharyya:2008mz} for general $d$).  
  An intriguing observation of \cite{VanRaamsdonk:2008fp} is the striking difference between the phenomenology of non-relativistic turbulent flows in 3+1 and 2+1 dimensions. In the 3+1 dimensional turbulent energy cascade, large scale eddies give rise to smaller scale eddies, eventually transferring energy down to scales where viscosity becomes important and energy is dissipated. In contrast, 2+1 dimensional turbulent flows are characterized by an inverse cascade, in which smaller scale eddies merge into large scale eddies, creating large long-lived vortical structures. If these qualitative differences extend to relativistic fluids, they would suggest a profound difference in gravitational dynamics in four and five dimensions. In particular, we might predict that black holes in AdS$_4$ would take much longer to equilibrate than AdS$_5$ black holes.  From the gravitational standpoint, this would certainly seem very surprising.

  More ambitiously, one may also consider fluids on curved manifolds (rather than just the Minkowski spacetime $\mathbb R^{d-1,1}$), as has been initiated in \cite{Bhattacharyya:2008ji}.   In addition, one can include matter in the bulk.  This allows for richer dynamics, but typically at the expense of losing universality.  Early examples of such extensions include considering the dilaton (which corresponds to forcing of the fluid) in \cite{Bhattacharyya:2008ji}, Maxwell $U(1)$ field \cite{Erdmenger:2008rm,Banerjee:2008th}, multiple Maxwell fields and scalars, magnetic and dyonic charges, as well as more exotic models (see e.g.\ \cite{Rangamani:2009xk} for further references).
Moreover, one can even extend the correspondence to non-conformal fluids \cite{Kanitscheider:2009as,David:2009np} as well as to non-relativistic fluids \cite{Rangamani:2008gi,Bhattacharyya:2008kq}, which allows us to make closer contact with familiar everyday systems.
Nevertheless, many future directions and puzzles remain, as well the need for further generalizations.  For example, of particular current interest is to understand the fluid/gravity correspondence for extremal fluids (and in particular superfluids) which are presently attracting much attention.  Also, to mimic many of the familiar aspects of fluid flows, we need to understand how to confine the fluid within walls in the gravity dual.  Still more ambitiously, to understand the rich phenomena rooted in quantum processes, we would like to get a better handle on stringy and quantum effects.


The fluid/gravity correspondence has a number of useful implications.
Among these is an improvement on the Israel-Stewart formalism \cite{Israel:1979wp}.
At first order, relativistic viscous fluid is described by a parabolic system, which leads to apparent causality violations.  The Israel-Stewart formalism renders the system hyperbolic by adding some 2nd order terms, but it does not capture all possible ones.  The fluid/gravity construction in effect prescribes the correct completion to render the full system causal.
Another intriguing consequence is the appearance of a new pseudo-vector
contribution to the charge current,\footnote{
Specifically, for Maxwell-Chern-Simons charged fluid, in addition to standard dissipative terms, a new  (parity-violating but CP
preserving) 1st order term appears in charge current:
$\ell^\mu = \epsilon_{\alpha \beta \gamma}^{\ \ \ \ \,  \mu} \, u^{\alpha} \nabla^{\beta} u^{\gamma}$.
} which  has been ignored by Landau\&Lifshitz \cite{Landau:1965uq}, but which may have potentially observable effects \cite{Son:2009tf}.
At a more general level, the framework of the fluid/gravity correspondence is rather reminiscent of the recently-developed `blackfold approach' to constructing higher-dimensional black holes \cite{Emparan:2009cs,Emparan:2009at}.  Whereas the former has an independent physical description of the bulk black hole provided by the dual conformal fluid, the latter is applicable to black holes with any asymptotics, as long as there is a separation of scales.
As a final example of useful application of the fluid/gravity correspondence, we note that the framework suggests a 
convenient rewriting of rotating AdS black holes in terms of fluid variables	\cite{Bhattacharyya:2008ji}, which is useful for analyzing its properties.


To summarize, one of the most intriguing features of the fluid/gravity correspondence is that it provides us with a window into the {\it generic} behavior of gravity in a nonlinear regime, mapping long-wavelength (but arbitrary amplitude) perturbations of AdS black holes to the more familiar physics of fluid dynamics.  Apart from the obvious conceptual advantages, one has a tremendous computational simplification for numerical studies of gravitational solutions since the fluid dynamics  lives in one lower dimension.
The boundary fluid stress tensor likewise contains new quantities of interest, namely the various transport coefficients which characterize the fluid.   This has been of interest in QCD phenomenology, especially in understanding certain characteristic features of the quark-gluon plasma. 
Hence the fluid/gravity correspondence provides a useful tool for not only
studying behavior of generic black holes in AdS, but also for geometrizing fluid dynamics  and for gaining insight into behavior of strongly coupled field theories, which exhibit similar features to certain real-world systems.  This is still a young and rapidly-expanding area of research, promising many further applications and fruitful developments.


Finally, let me close by revisiting the subtitle of my talk, so as to specify the new perspective on the black hole membrane paradigm that the fluid/gravity correspondence offers  \cite{Hubeny:2009zz}.  As described earlier, the conventional membrane paradigm \cite{Thorne:1986iy} provides a simple picture of black hole dynamics in terms of a fluid living on a `membrane' (the stretched horizon) just outside the event horizon.   Stepping back to take a more general view of trying to encode the black hole dynamics by fluid dynamics localized on a membrane in the spacetime, a natural question is where should such a membrane live?  Perhaps the most obvious candidate is the event horizon; but this is problematic due to its null nature, and more importantly, because it is defined globally, requiring the knowledge of full future evolution of the spacetime.  Alternately, several (quasi)local notions of a black hole have been proposed, such as the so-called dynamical horizon \cite{Booth:2005qc,Ashtekar:2004cn}, which however are spacelike surfaces inside the event horizon, and therefore do not admit the standard notion of evolution.  A more popular suggestion is the stretched horizon, which is the formulation given by the membrane paradigm \cite{Thorne:1986iy}.  However, there likewise remain ambiguities in localizing  the stretched horizon. 
Within the fluid/gravity correspondence, the entire spacetime evolution is mapped to the dynamics of a conformal fluid, which, albeit reminiscent of the membrane paradigm, has one important twist: the membrane lives on the {\it boundary} of the spacetime (which is unambiguously defined and admits a fluid description with well-defined dynamics), and gives a perfect mirror of the full bulk physics. This  ``membrane at the end of the universe" picture is a natural consequence of the holographic nature of the fluid/gravity correspondence.

\subsection*{Acknowledgements}

It is a pleasure to thank my collaborators, Sayantani Bhattacharyya, R.\ Loganayagam,  Gautam Mandal, Shiraz Minwalla, Takeshi Morita, Mukund Rangamani, Harvey Reall, and Mark Van Raamsdonk for wonderful collaborations on various aspects discussed in this review.
VH is partly supported by STFC Rolling grant.


\begin{thebibliography}{10}

\bibitem{Bekenstein:1973ur}
J.~D. Bekenstein, ``{Black holes and entropy},''
\href{http://dx.doi.org/10.1103/PhysRevD.7.2333}{{\em Phys. Rev.} {\bf D7}
  (1973)  2333--2346}.

\bibitem{Hawking:1974sw}
S.~W. Hawking, ``{Particle Creation by Black Holes},''
\href{http://dx.doi.org/10.1007/BF02345020}{{\em Commun. Math. Phys.} {\bf 43}
  (1975)  199--220}.

\bibitem{Bardeen:1973gs}
J.~M. Bardeen, B.~Carter, and S.~W. Hawking, ``{The Four laws of black hole
  mechanics},''
\href{http://dx.doi.org/10.1007/BF01645742}{{\em Commun. Math. Phys.} {\bf 31}
  (1973)  161--170}.

\bibitem{Unruh:1980cg}
W.~G. Unruh, ``{Experimental black hole evaporation},''
\href{http://dx.doi.org/10.1103/PhysRevLett.46.1351}{{\em Phys. Rev. Lett.}
  {\bf 46} (1981)  1351--1353}.

\bibitem{Cardoso:2006ks}
V.~Cardoso and O.~J.~C. Dias, ``{Gregory-Laflamme and Rayleigh-Plateau
  instabilities},'' \href{http://dx.doi.org/10.1103/PhysRevLett.96.181601}{{\em
  Phys. Rev. Lett.} {\bf 96} (2006)  181601},
\href{http://arxiv.org/abs/hep-th/0602017}{{\tt arXiv:hep-th/0602017}}.

\bibitem{Gregory:1993vy}
R.~Gregory and R.~Laflamme, ``{Black strings and p-branes are unstable},''
  \href{http://dx.doi.org/10.1103/PhysRevLett.70.2837}{{\em Phys. Rev. Lett.}
  {\bf 70} (1993)  2837--2840},
\href{http://arxiv.org/abs/hep-th/9301052}{{\tt arXiv:hep-th/9301052}}.

\bibitem{Plateau:1873fk}
J.~Plateau, {\em Statique Experimentale et Theorique des Liquides Soumis aux
  Seules Forces Moleculaires}.
\newblock Gauthier-Villars, 1873.

\bibitem{Rayleigh:1878uq}
L.~Rayleigh {\em Proc. Lond. Math. Soc.} {\bf 10} (1878)  4.

\bibitem{Rezzolla:2010df}
  L.~Rezzolla, R.~P.~Macedo and J.~L.~Jaramillo,
  ``{Understanding the `anti-kick' in the merger of binary black holes},''
 \href{http://dx.doi.org/10.1103/PhysRevLett.104.221101}{{\em Phys.\ Rev.\ Lett.}  {\bf 104} (2010) 221101},
 \href{http://arxiv.org/abs/1003.0873}{{\tt arXiv:1003.0873 [gr-qc]}}.

\bibitem{Thorne:1986iy}
K.~S. Thorne, R.~H. Price, and D.~A. MacDonald, {\em Black Holes: The Membrane
  Paradigm}.
\newblock Yale University Press, New Haven, 1986.

\bibitem{Damour:1978cg}
T.~Damour, ``{Black Hole Eddy Currents},''
\href{http://dx.doi.org/10.1103/PhysRevD.18.3598}{{\em Phys. Rev.} {\bf D18}
  (1978)  3598--3604}.

\bibitem{Bhattacharyya:2008jc}
S.~Bhattacharyya, V.~E. Hubeny, S.~Minwalla, and M.~Rangamani, ``{Nonlinear
  Fluid Dynamics from Gravity},''
  \href{http://dx.doi.org/10.1088/1126-6708/2008/02/045}{{\em JHEP} {\bf 02}
  (2008)  045},
\href{http://arxiv.org/abs/0712.2456}{{\tt arXiv:0712.2456 [hep-th]}}.

\bibitem{Policastro:2001yc}
G.~Policastro, D.~T. Son, and A.~O. Starinets, ``{The shear viscosity of
  strongly coupled N = 4 supersymmetric Yang-Mills plasma},''
  \href{http://dx.doi.org/10.1103/PhysRevLett.87.081601}{{\em Phys. Rev. Lett.}
  {\bf 87} (2001)  081601},
\href{http://arxiv.org/abs/hep-th/0104066}{{\tt arXiv:hep-th/0104066}}.

\bibitem{Janik:2005zt}
R.~A. Janik and R.~B. Peschanski, ``{Asymptotic perfect fluid dynamics as a
  consequence of AdS/CFT},''
  \href{http://dx.doi.org/10.1103/PhysRevD.73.045013}{{\em Phys. Rev.} {\bf
  D73} (2006)  045013},
\href{http://arxiv.org/abs/hep-th/0512162}{{\tt arXiv:hep-th/0512162}}.

\bibitem{Bhattacharyya:2007vs}
S.~Bhattacharyya, S.~Lahiri, R.~Loganayagam, and S.~Minwalla, ``{Large rotating
  AdS black holes from fluid mechanics},''
  \href{http://dx.doi.org/10.1088/1126-6708/2008/09/054}{{\em JHEP} {\bf 09}
  (2008)  054},
\href{http://arxiv.org/abs/0708.1770}{{\tt arXiv:0708.1770 [hep-th]}}.

\bibitem{Rangamani:2009xk}
M.~Rangamani, ``{Gravity \& Hydrodynamics: Lectures on the fluid-gravity
  correspondence},''
  \href{http://dx.doi.org/10.1088/0264-9381/26/22/224003}{{\em Class. Quant.
  Grav.} {\bf 26} (2009)  224003},
\href{http://arxiv.org/abs/0905.4352}{{\tt arXiv:0905.4352 [hep-th]}}.

\bibitem{Hubeny:2010fk}
V.~E. Hubeny, ``Fluid dynamics from gravity,'' {\em From Gravity to Thermal Gauge Theories},
  Lecture Notes in Physics {\bf 828}, Springer (2011).
  \url{http://www.springer.com/physics/book/978-3-642-04863-0}.

\bibitem{Hubeny:2010ry}
V.~E. Hubeny and M.~Rangamani, ``{A Holographic view on physics out of
  equilibrium},'' \href{http://arxiv.org/abs/1006.3675}{{\tt arXiv:1006.3675
  [hep-th]}}.

\bibitem{Mateos:2007ay}
D.~Mateos, ``{String Theory and Quantum Chromodynamics},''
  \href{http://dx.doi.org/10.1088/0264-9381/24/21/S01}{{\em Class. Quant.
  Grav.} {\bf 24} (2007)  S713--S740},
\href{http://arxiv.org/abs/0709.1523}{{\tt arXiv:0709.1523 [hep-th]}}.

\bibitem{Gubser:2009md}
S.~S. Gubser and A.~Karch, ``{From gauge-string duality to strong interactions:
  a Pedestrian's Guide},''
  \href{http://dx.doi.org/10.1146/annurev.nucl.010909.083602}{{\em Ann. Rev.
  Nucl. Part. Sci.} {\bf 59} (2009)  145--168},
\href{http://arxiv.org/abs/0901.0935}{{\tt arXiv:0901.0935 [hep-th]}}.

\bibitem{McGreevy:2009xe}
J.~McGreevy, ``{Holographic duality with a view toward many-body physics},''
\href{http://arxiv.org/abs/0909.0518}{{\tt arXiv:0909.0518 [hep-th]}}.

\bibitem{Hartnoll:2009sz}
S.~A. Hartnoll, ``{Lectures on holographic methods for condensed matter
  physics},'' \href{http://dx.doi.org/10.1088/0264-9381/26/22/224002}{{\em
  Class. Quant. Grav.} {\bf 26} (2009)  224002},
\href{http://arxiv.org/abs/0903.3246}{{\tt arXiv:0903.3246 [hep-th]}}.

\bibitem{Fefferman:2000fk}
C.~Fefferman, ``{Existence and smoothness of the Navier-Stokes equation},''
  {\em Clay Millenium Problems} (2000).

\bibitem{Bhattacharyya:2008xc}
S.~Bhattacharyya, V.~E. Hubeny, R.~Loganayagam, G.~Mandal, S.~Minwalla,
  T.~Morita, M.~Rangamani, and H.~S. Reall, ``{Local Fluid Dynamical Entropy
  from Gravity},'' {\em JHEP} {\bf 06} (2008)  055,
\href{http://arxiv.org/abs/0803.2526}{{\tt arXiv:0803.2526 [hep-th]}}.

\bibitem{Bhattacharyya:2008mz}
S.~Bhattacharyya, R.~Loganayagam, I.~Mandal, S.~Minwalla, and A.~Sharma,
  ``{Conformal Nonlinear Fluid Dynamics from Gravity in Arbitrary
  Dimensions},'' \href{http://dx.doi.org/10.1088/1126-6708/2008/12/116}{{\em
  JHEP} {\bf 12} (2008)  116},
\href{http://arxiv.org/abs/0809.4272}{{\tt arXiv:0809.4272 [hep-th]}}.

\bibitem{Maldacena:1997re}
J.~M. Maldacena, ``{The large N limit of superconformal field theories and
  supergravity},'' {\em Adv. Theor. Math. Phys.} {\bf 2} (1998)  231--252,
\href{http://arxiv.org/abs/hep-th/9711200}{{\tt arXiv:hep-th/9711200}}.

\bibitem{Gubser:1998bc}
S.~S. Gubser, I.~R. Klebanov, and A.~M. Polyakov, ``{Gauge theory correlators
  from non-critical string theory},''
  \href{http://dx.doi.org/10.1016/S0370-2693(98)00377-3}{{\em Phys. Lett.} {\bf
  B428} (1998)  105--114},
\href{http://arxiv.org/abs/hep-th/9802109}{{\tt arXiv:hep-th/9802109}}.

\bibitem{Witten:1998qj}
E.~Witten, ``{Anti-de Sitter space and holography},'' {\em Adv. Theor. Math.
  Phys.} {\bf 2} (1998)  253--291,
\href{http://arxiv.org/abs/hep-th/9802150}{{\tt arXiv:hep-th/9802150}}.

\bibitem{Aharony:1999ti}
O.~Aharony, S.~S. Gubser, J.~M. Maldacena, H.~Ooguri, and Y.~Oz, ``{Large N
  field theories, string theory and gravity},''
  \href{http://dx.doi.org/10.1016/S0370-1573(99)00083-6}{{\em Phys. Rept.} {\bf
  323} (2000)  183--386},
\href{http://arxiv.org/abs/hep-th/9905111}{{\tt arXiv:hep-th/9905111}}.

\bibitem{DHoker:2002aw}
E.~D'Hoker and D.~Z. Freedman, ``{Supersymmetric gauge theories and the AdS/CFT
  correspondence},''
\href{http://arxiv.org/abs/hep-th/0201253}{{\tt arXiv:hep-th/0201253}}.

\bibitem{Polchinski:2010hw}
J.~Polchinski, ``{Introduction to Gauge/Gravity Duality},''
  \href{http://arxiv.org/abs/1010.6134}{{\tt arXiv:1010.6134 [hep-th]}}.

\bibitem{tHooft:1993gx}
G.~'t~Hooft, ``{Dimensional reduction in quantum gravity},''
\href{http://arxiv.org/abs/gr-qc/9310026}{{\tt arXiv:gr-qc/9310026}}.

\bibitem{Susskind:1994vu}
L.~Susskind, ``{The World as a hologram},''
  \href{http://dx.doi.org/10.1063/1.531249}{{\em J. Math. Phys.} {\bf 36}
  (1995)  6377--6396},
\href{http://arxiv.org/abs/hep-th/9409089}{{\tt arXiv:hep-th/9409089}}.

\bibitem{Balasubramanian:1999re}
V.~Balasubramanian and P.~Kraus, ``{A stress tensor for anti-de Sitter
  gravity},'' \href{http://dx.doi.org/10.1007/s002200050764}{{\em Commun. Math.
  Phys.} {\bf 208} (1999)  413--428},
\href{http://arxiv.org/abs/hep-th/9902121}{{\tt arXiv:hep-th/9902121}}.

\bibitem{deHaro:2000xn}
S.~de~Haro, S.~N. Solodukhin, and K.~Skenderis, ``{Holographic reconstruction
  of spacetime and renormalization in the AdS/CFT correspondence},''
  \href{http://dx.doi.org/10.1007/s002200100381}{{\em Commun. Math. Phys.} {\bf
  217} (2001)  595--622},
\href{http://arxiv.org/abs/hep-th/0002230}{{\tt arXiv:hep-th/0002230}}.

\bibitem{Landau:1965uq}
L.~D. Landau and E.~M. Lifshitz, {\em {Fluid Mechanics (Course of Theoretical
  Physics, Vol. 6)}}.
\newblock Butterworth-Heinemann, 1965.

\bibitem{Loganayagam:2008is}
R.~Loganayagam, ``{Entropy Current in Conformal Hydrodynamics},''
  \href{http://dx.doi.org/10.1088/1126-6708/2008/05/087}{{\em JHEP} {\bf 05}
  (2008)  087},
\href{http://arxiv.org/abs/0801.3701}{{\tt arXiv:0801.3701 [hep-th]}}.

\bibitem{Kokkotas:1999bd}
K.~D. Kokkotas and B.~G. Schmidt, ``{Quasi-normal modes of stars and black
  holes},'' {\em Living Rev. Rel.} {\bf 2} (1999)  2,
\href{http://arxiv.org/abs/gr-qc/9909058}{{\tt arXiv:gr-qc/9909058}}.

\bibitem{Horowitz:1999jd}
G.~T. Horowitz and V.~E. Hubeny, ``{Quasinormal modes of AdS black holes and
  the approach to thermal equilibrium},''
  \href{http://dx.doi.org/10.1103/PhysRevD.62.024027}{{\em Phys. Rev.} {\bf
  D62} (2000)  024027},
\href{http://arxiv.org/abs/hep-th/9909056}{{\tt arXiv:hep-th/9909056}}.

\bibitem{Berti:2009kk}
E.~Berti, V.~Cardoso, and A.~O. Starinets, ``{Quasinormal modes of black holes
  and black branes},''
  \href{http://dx.doi.org/10.1088/0264-9381/26/16/163001}{{\em Class. Quant.
  Grav.} {\bf 26} (2009)  163001},
\href{http://arxiv.org/abs/0905.2975}{{\tt arXiv:0905.2975 [gr-qc]}}.

\bibitem{Policastro:2002se}
G.~Policastro, D.~T. Son, and A.~O. Starinets, ``{From AdS/CFT correspondence
  to hydrodynamics},'' {\em JHEP} {\bf 09} (2002)  043,
\href{http://arxiv.org/abs/hep-th/0205052}{{\tt arXiv:hep-th/0205052}}.

\bibitem{Policastro:2002tn}
G.~Policastro, D.~T. Son, and A.~O. Starinets, ``{From AdS/CFT correspondence
  to hydrodynamics. II: Sound waves},'' {\em JHEP} {\bf 12} (2002)  054,
\href{http://arxiv.org/abs/hep-th/0210220}{{\tt arXiv:hep-th/0210220}}.

\bibitem{Son:2007vk}
D.~T. Son and A.~O. Starinets, ``{Viscosity, Black Holes, and Quantum Field
  Theory},''
  \href{http://dx.doi.org/10.1146/annurev.nucl.57.090506.123120}{{\em Ann. Rev.
  Nucl. Part. Sci.} {\bf 57} (2007)  95--118},
\href{http://arxiv.org/abs/0704.0240}{{\tt arXiv:0704.0240 [hep-th]}}.

\bibitem{Morgan:2009pn}
J.~Morgan, V.~Cardoso, A.~S. Miranda, C.~Molina, and V.~T. Zanchin,
  ``{Gravitational quasinormal modes of AdS black branes in d spacetime
  dimensions},'' \href{http://dx.doi.org/10.1088/1126-6708/2009/09/117}{{\em
  JHEP} {\bf 09} (2009)  117},
\href{http://arxiv.org/abs/0907.5011}{{\tt arXiv:0907.5011 [hep-th]}}.

\bibitem{Kovtun:2004de}
P.~Kovtun, D.~T. Son, and A.~O. Starinets, ``{Viscosity in strongly interacting
  quantum field theories from black hole physics},''
  \href{http://dx.doi.org/10.1103/PhysRevLett.94.111601}{{\em Phys. Rev. Lett.}
  {\bf 94} (2005)  111601},
\href{http://arxiv.org/abs/hep-th/0405231}{{\tt arXiv:hep-th/0405231}}.

\bibitem{Schafer:2009dj}
T.~Schafer and D.~Teaney, ``{Nearly Perfect Fluidity: From Cold Atomic Gases to
  Hot Quark Gluon Plasmas},''
  \href{http://dx.doi.org/10.1088/0034-4885/72/12/126001}{{\em Rept. Prog.
  Phys.} {\bf 72} (2009)  126001},
\href{http://arxiv.org/abs/0904.3107}{{\tt arXiv:0904.3107 [hep-ph]}}.

\bibitem{Baier:2007ix}
R.~Baier, P.~Romatschke, D.~T. Son, A.~O. Starinets, and M.~A. Stephanov,
  ``{Relativistic viscous hydrodynamics, conformal invariance, and
  holography},'' \href{http://dx.doi.org/10.1088/1126-6708/2008/04/100}{{\em
  JHEP} {\bf 04} (2008)  100},
\href{http://arxiv.org/abs/0712.2451}{{\tt arXiv:0712.2451 [hep-th]}}.

\bibitem{Haack:2008cp}
M.~Haack and A.~Yarom, ``{Nonlinear viscous hydrodynamics in various dimensions
  using AdS/CFT},'' \href{http://dx.doi.org/10.1088/1126-6708/2008/10/063}{{\em
  JHEP} {\bf 10} (2008)  063},
\href{http://arxiv.org/abs/0806.4602}{{\tt arXiv:0806.4602 [hep-th]}}.

\bibitem{VanRaamsdonk:2008fp}
M.~Van~Raamsdonk, ``{Black Hole Dynamics From Atmospheric Science},''
  \href{http://dx.doi.org/10.1088/1126-6708/2008/05/106}{{\em JHEP} {\bf 05}
  (2008)  106},
\href{http://arxiv.org/abs/0802.3224}{{\tt arXiv:0802.3224 [hep-th]}}.

\bibitem{Bhattacharyya:2008ji}
S.~Bhattacharyya, R.~Loganayagam, S.~Minwalla, S.~Nampuri, S.~P. Trivedi, and
  S.~R. Wadia, ``{Forced Fluid Dynamics from Gravity},''
  \href{http://dx.doi.org/10.1088/1126-6708/2009/02/018}{{\em JHEP} {\bf 02}
  (2009)  018},
\href{http://arxiv.org/abs/0806.0006}{{\tt arXiv:0806.0006 [hep-th]}}.

\bibitem{Erdmenger:2008rm}
J.~Erdmenger, M.~Haack, M.~Kaminski, and A.~Yarom, ``{Fluid dynamics of
  R-charged black holes},''
  \href{http://dx.doi.org/10.1088/1126-6708/2009/01/055}{{\em JHEP} {\bf 01}
  (2009)  055},
\href{http://arxiv.org/abs/0809.2488}{{\tt arXiv:0809.2488 [hep-th]}}.

\bibitem{Banerjee:2008th}
N.~Banerjee, J.~Bhattacharya, S.~Bhattacharyya, S.~Dutta, R.~Loganayagam, and
  P.~Surowka, ``{Hydrodynamics from charged black branes},''
\href{http://arxiv.org/abs/0809.2596}{{\tt arXiv:0809.2596 [hep-th]}}.

\bibitem{Kanitscheider:2009as}
I.~Kanitscheider and K.~Skenderis, ``{Universal hydrodynamics of non-conformal
  branes},'' \href{http://dx.doi.org/10.1088/1126-6708/2009/04/062}{{\em JHEP}
  {\bf 04} (2009)  062},
\href{http://arxiv.org/abs/0901.1487}{{\tt arXiv:0901.1487 [hep-th]}}.

\bibitem{David:2009np}
J.~R. David, M.~Mahato, and S.~R. Wadia, ``{Hydrodynamics from the D1-brane},''
  \href{http://dx.doi.org/10.1088/1126-6708/2009/04/042}{{\em JHEP} {\bf 04}
  (2009)  042},
\href{http://arxiv.org/abs/0901.2013}{{\tt arXiv:0901.2013 [hep-th]}}.

\bibitem{Rangamani:2008gi}
M.~Rangamani, S.~F. Ross, D.~T. Son, and E.~G. Thompson, ``{Conformal
  non-relativistic hydrodynamics from gravity},''
  \href{http://dx.doi.org/10.1088/1126-6708/2009/01/075}{{\em JHEP} {\bf 01}
  (2009)  075},
\href{http://arxiv.org/abs/0811.2049}{{\tt arXiv:0811.2049 [hep-th]}}.

\bibitem{Bhattacharyya:2008kq}
S.~Bhattacharyya, S.~Minwalla, and S.~R. Wadia, ``{The Incompressible
  Non-Relativistic Navier-Stokes Equation from Gravity},''
  \href{http://dx.doi.org/10.1088/1126-6708/2009/08/059}{{\em JHEP} {\bf 08}
  (2009)  059},
\href{http://arxiv.org/abs/0810.1545}{{\tt arXiv:0810.1545 [hep-th]}}.

\bibitem{Israel:1979wp}
W.~Israel and J.~M. Stewart, ``{Transient relativistic thermodynamics and
  kinetic theory},''
\href{http://dx.doi.org/10.1016/0003-4916(79)90130-1}{{\em Ann. Phys.} {\bf
  118} (1979)  341--372}.

\bibitem{Son:2009tf}
D.~T. Son and P.~Surowka, ``{Hydrodynamics with Triangle Anomalies},''
  \href{http://dx.doi.org/10.1103/PhysRevLett.103.191601}{{\em Phys.Rev.Lett.}
  {\bf 103} (2009)  191601}, \href{http://arxiv.org/abs/0906.5044}{{\tt
  arXiv:0906.5044 [hep-th]}}.

\bibitem{Emparan:2009cs}
R.~Emparan, T.~Harmark, V.~Niarchos, and N.~A. Obers, ``{Blackfolds},''
  \href{http://dx.doi.org/10.1103/PhysRevLett.102.191301}{{\em Phys. Rev.
  Lett.} {\bf 102} (2009)  191301},
\href{http://arxiv.org/abs/0902.0427}{{\tt arXiv:0902.0427 [hep-th]}}.

\bibitem{Emparan:2009at}
R.~Emparan, T.~Harmark, V.~Niarchos, and N.~A. Obers, ``{Essentials of
  Blackfold Dynamics},''
\href{http://arxiv.org/abs/0910.1601}{{\tt arXiv:0910.1601 [Unknown]}}.

\bibitem{Hubeny:2009zz}
V.~E. Hubeny, M.~Rangamani, S.~Minwalla, and M.~Van~Raamsdonk, ``{The
  fluid-gravity correspondence: The membrane at the end of the universe},''
\href{http://dx.doi.org/10.1142/S0218271808014084}{{\em Int. J. Mod. Phys.}
  {\bf D17} (2009)  2571--2576}.

\bibitem{Booth:2005qc}
I.~Booth, ``{Black hole boundaries},''
  \href{http://dx.doi.org/10.1139/p05-063}{{\em Can. J. Phys.} {\bf 83} (2005)
  1073--1099},
\href{http://arxiv.org/abs/gr-qc/0508107}{{\tt arXiv:gr-qc/0508107}}.

\bibitem{Ashtekar:2004cn}
A.~Ashtekar and B.~Krishnan, ``{Isolated and dynamical horizons and their
  applications},'' {\em Living Rev. Rel.} {\bf 7} (2004)  10,
\href{http://arxiv.org/abs/gr-qc/0407042}{{\tt arXiv:gr-qc/0407042}}.

\end{thebibliography}

\providecommand{\href}[2]{#2}\begingroup\raggedright\endgroup

\end{document}